\documentclass[11pt]{iopart}
\usepackage{iopams,amssymb}
\usepackage{graphicx}
\usepackage{dcolumn}
\usepackage{bm}
\usepackage{epsfig}

\begin{document}

\title[TBA Phase diagram of the strong coupling ladder compounds]{Note on 
the thermodynamic Bethe Ansatz approach to the quantum 
phase diagram of the strong coupling ladder compounds}

\author{M.T. Batchelor$^{\dag}$, X.-W. Guan$^{\dag}$, A. Foerster$^{\ddag}$ 
and  H.-Q. Zhou$^{\ddag \ddag}$}

\address{$^{\dag}$ Department of Theoretical Physics, Research School of Physical 
Sciences and Engineering \& Centre for Mathematics and its Applications,
Mathematical Sciences Institute, Australian National University, 
Canberra ACT 0200, Australia}

\address{$^{\ddag}$ Instituto de F\'{\i}sica da UFRGS,
Av. Bento Gon\c{c}alves 9500, Porto Alegre, RS - Brazil}

\address{$^{\ddag \ddag}$ Centre for Mathematical Physics, 
University of Queensland, Queensland 4072, Australia}

\date{\today}

\begin{abstract}
We investigate the low-temperature phase diagram of the exactly solved
$su(4)$ two-leg spin ladder as a function of the rung coupling $J_{\perp}$
and magnetic field $H$ by means of the thermodynamic Bethe Ansatz (TBA).
In the absence of a magnetic field the model exhibits
three quantum phases, while in the presence of a strong magnetic field
there is no singlet ground state for ferromagnetic rung coupling. 
For antiferromagnetic rung coupling, there is a gapped phase in the 
regime $H < H_{c1}$, a fully polarized gapped phase for $H > H_{c2}$ 
and a Luttinger liquid magnetic phase in the regime $H_{c1} < H < H_{c2}$.
The critical behaviour derived using the TBA is consistent with the existing 
experimental, numerical and perturbative results for the strong coupling ladder  
compounds.
This includes the spin excitation gap and the critical fields $H_{c1}$ 
and $H_{c2}$, which are in excellent agreement with the experimental values 
for the known strong coupling ladder compounds 
($5$IAP)$_2$CuBr$_4$$\cdot$$2$H$_2$O, Cu$_{2}$(C$_5$H$_{12}$N$_2$)$_2$Cl$_4$ and  
(C$_5$H$_{12}$N)$_2$CuBr$_{4}$. 
In addition we predict the spin gap $\Delta \approx J_{\perp}-\frac{1}{2}J_{\parallel}$ 
for the weak coupling compounds with $ J_{\perp} \sim J_{\parallel}$, such as 
(VO)$_2$P$_2$O$_7$, and also show that the gap opens for arbitrary 
$J_{\perp}/ J_{\parallel}$.

\end{abstract}

\maketitle

Recently there has been considerable theoretical and experimental interest in 
spin ladder systems. With the rapid progress presently being made in
nano-engineering, many compounds with a ladder structure have been 
experimentally realized,
such as SrCu$_2$(BO$_3$)$_2$, Cu$_2$(C$_5$H$_{12}$N$_2$)$_2$ Cl$_4$,
(C$_5$H$_{12}$N)$_2$CuBr$_{4}$ and KCuCl$_3$ \cite{exp1}.
The existence of a spin gap, magnetization plateaus, superconductivity 
under hole doping, etc,  are examples of some interesting physical properties 
that may be observed in experiments involving ladder compounds (see, e.g.,
Refs \cite{exp1}-\cite{exp7} and references therein).
From the theoretical point of view, most of the results for ladder systems
were initially obtained from studies of the standard Heisenberg ladder, which 
in contrast to its one-dimensional counterpart, cannot be solved exactly. 
Subsequently, other generalised ladder models have been proposed \cite{ladd1} and 
analysed through various numerical, approximate and exact approaches 
\cite{numer1,numer2,FT1,FT2}.

On the other hand, although some exactly solved or integrable ladder models have 
been introduced \cite{ladd4,ladd3,ladd5}, 
none have been used to predict physical properties which
could be compared directly with experimental data, such as the critical magnetic fields.
In this context, the integrable spin ladder model based on the $su(4)$ 
algebra \cite{ladd4}
appears to be a good candidate for this purpose, since its 
Hamiltonian consists of the standard Heisenberg ladder model with an extra biquadratic spin 
interaction term along the legs, the physical importance of which has been noted \cite{ladd1}.
In the strong coupling limit, the contribution 
to the low-temperature physics from the biquadratic term 
is minimal, and as a consequence, the model exhibits similar critical behavior to the 
standard Heisenberg ladder.
Therefore it is reasonable to expect that the integrable $su(4)$ ladder model 
can well describe the low-temperature critical behavior of the strong coupling
ladder materials. In addition, by properly minimizing the intrachain coupling 
in the integrable ladder Hamiltonian, the model may also be used to describe the weak
coupling compounds.

Here we show that this is in fact true in the strong coupling regime by 
investigating the quantum phase diagram of the integrable $su(4)$ ladder,
which can be tested by experiments.
Our analytic expression for the gap, $\Delta =J_{\perp}-4J_{\parallel}/\gamma$, 
and the critical fields, $\mu_BgH_{c1}=\Delta$ and 
$\mu_BgH_{c2}= J_{\perp}+4J_{\parallel}/\gamma$, where $\gamma$ is a
rescaling constant, can be applied in general to strong coupling ladder compounds 
with Heisenberg interactions, 
such as   ($5$IAP)$_2$CuBr$_4 \cdot 2$H$_2$O, Cu$_2$(C$_5$H$_{12}$N$_2$)$_2$Cl$_4$,
(C$_5$H$_{12}$N)$_2$CuBr$_{4}$ and KCuCl$_3$,  by choosing $\gamma \approx 4$. 
For weak ($J_{\perp} \sim J_{\parallel}$) coupling compounds, such as $(VO)_2P_2O7$,
the choice $\gamma \approx 8$ determines a good fit for the gap \cite{exp1,weak}.
In addition, in the presence of a strong magnetic field, we shown that the 
quantum phase diagram and the critical behavior predicted from the thermodynamic
Bethe Ansatz (TBA) are in a good agreement with the experimental results for the 
above-mentioned compounds. 
We also show that the gap opens for an arbitrary value of
$J_{\perp}/ J_{\parallel}$, in accordance with the experimental results.

\vskip 3mm
\noindent
{\bf I. The Model.} 
We consider the phase diagram of the simplest integrable spin ladder \cite{ladd4}
\begin{equation}
H=\frac{J_{\parallel}}{\gamma}H_{{\rm leg}}+J_{\perp}\sum_{j=1}^{L}\vec{S}_j\vec{T}_j+
h\sum_{j=1}^{L}(S_j^z+T^z_j),\label{Ham}
\end{equation}
where
\begin{equation}
H_{{\rm
leg}}=\sum_{j=1}^{L}\left(\vec{S}_j\vec{S}_{j+1}+\vec{T}_j\vec{T}_{j+1}+
4\vec{S}_j\vec{S}_{j+1}\vec{T}_j\vec{T}_{j+1}\right).
\label{intra} 
\end{equation}
Here $\vec{S }_j$ and $\vec{T}_j$ are the
standard spin-$\frac12$ operators acting on site $j$ of the upper and lower
legs, respectively, $J_{\parallel}$ and $J_{\perp}$ are the intra (leg) 
and interchain (rung) couplings and $h$ is the magnetic field.  Throughout,
$L$ is the number of rungs and periodic boundary conditions are imposed. 
Essentially, the competition between the rung and leg
couplings and the magnetic field $h$ determines the
physical properties and the critical behavior of the system.
In order to facilitate the comparsion with real compounds, the intrachain part of this 
model (\ref{intra}) can be minimized through a rescaling constant $\gamma$.
In comparison with the standard spin-$\frac12$ Heisenberg ladder \cite{exp1,exp2,exp4,Hayw}, 
the above Hamiltonian contains a four-spin interaction term, which
minimizes the Haldane phase \cite{ladd1} and causes a shift of the
critical value of the rung coupling $J_{\perp}$ at which the model
becomes massive.  It is well established that this Hamiltonian
is integrable and its leg part $H_{{\rm leg}}$ is (up to a constant) simply
the permutation operator corresponding to the $su(4)$ algebra
\cite{ladd4}.  In addition, after the convenient change of basis,
$|1\rangle = \frac{1}{\sqrt{2}}\left(|\uparrow \downarrow\rangle
-|\downarrow \uparrow \rangle \right),\, |2\rangle
=|\uparrow\uparrow\rangle,\, |3\rangle =
\frac{1}{\sqrt{2}}\left(|\uparrow \downarrow\rangle +|\downarrow
\uparrow \rangle \right),\, |4\rangle =|\downarrow\downarrow\rangle$,
where the first state denotes the rung singlet and the three others the 
components of the triplet, the leg part remains of the same form while the
rung term becomes diagonal. 
This rung term reduces the $su(4)$ symmetry of $H_{{\rm leg}}$ to $su(3)\oplus u(1)$
symmetry. Switching on the magnetic field breaks this symmetry further due to  
Zeeman splitting. This Hamiltonian can be diagonalized using the nested algebraic 
Bethe Ansatz (BA) with three levels. 
It is worth noting that for the ladder Hamiltonian (\ref{Ham}), the singlet rung state 
is energetically favoured for $J_\perp>0$, whereas the triplet 
rung state is favoured for $J_\perp<0$. 
On applying the magnetic field, component $|2 \rangle$ of the triplet 
is energetically favoured. We will use these properties to our
advantage by doing calculations with the choice of ordering for which the 
BA reference state is the closest to the true groundstate of the system.

The underlying BA equations for Hamiltonian (\ref{Ham}) are well known \cite{BA} and
consist of a set of three coupled equations depending on the
flavors, $v$, $u$ and $w$. Adopting the string conjecture \cite{TBA,Lee} and
taking the thermodynamic limit, 
the densities of the three flavors, $\rho ^{(1)}_n(v)$, $\rho^{(2)}_n(u)$ and 
$\rho ^{(3)}_n(w)$, can be defined as usual. After some manipulations, 
the BA equations reduce to
\begin{equation}
\left(\begin{array}{c}
\ln(1+\eta _n^{(1)})\\
\ln(1+\eta _n^{(2)})\\
\ln(1+\eta _n^{(3)})
\end{array}\right)
=\frac{G}{T}+K*
\left(\begin{array}{l}
\ln(1+{\eta _m^{(1)}}^{-1})\\
\ln(1+{\eta _m^{(2)}}^{-1})\\
\ln(1+{\eta _m^{(3)}}^{-1})
\end{array}\right),\label{TBA}
\end{equation}
where $\rho^{(1)h}_n(v)$, $\rho ^{(2)h}_n(u)$ and $\rho ^{(3)h}_n(w)$ denote the
hole densities and
\begin{eqnarray}
& &K =
\left(\begin{array}{ccc}
\sum_m A_{nm}&-\sum_m a_{nm}&0\\
-\sum_m a_{nm}&\sum_m A_{nm}&-\sum_m a_{nm}\\
0&-\sum_m a_{nm}&\sum _{m}A_{nm}
\end{array}\right),
\end{eqnarray}
where
\begin{eqnarray}
A_{nm}(\lambda)&=&\delta(\lambda)\delta_{nm}+(1-\delta_{nm})a_{|n-m|}(\lambda)
+a_{n+m}(\lambda)\nonumber\\
& & \hspace{2cm}+2\sum^{{\rm Min}(n,m)-1}_{l=1}a_{|n-m|+2l}(\lambda),\\
a_{nm}(\lambda)&=&\sum^{{\rm Min}(n,m)}_{l=1}a_{n+m+1-2l}(\lambda),
\end{eqnarray}
with $a_n(\lambda)=\frac{1}{2\pi}\frac{n}{n^2/4+\lambda ^2}$. 
The symbol $*$ denotes convolution and 
$\eta ^{(l)}_n(\lambda)=\rho^{(l)h}_n(\lambda)/\rho ^{(l)}_n(\lambda):=
\exp(\epsilon^{(l)}_n(\lambda)/T),\, l=1,2,3$. The dressed energy
$\epsilon^{(l)}_n$ plays the role of an excitation energy measured
from the Fermi level.  The driving matrix  $G$ depends on the choice of
the reference state. Explicitly, for $J_{\perp}<0$, 
$G={\rm column}(-\frac{J_{\parallel}}{\gamma} 2\pi a_n+nh,nh,-n(J_{\perp}+h))$ 
giving the free energy
\begin{equation}
\frac{F(T,h)}{L}=-h-T\int_{-\infty}^{\infty}\sum_{n=1}^{\infty}a_n(\lambda)
\ln(1+e^{-\frac{\epsilon^{(1)}_n(\lambda)}{T}})d\lambda. \label{FE}
\end{equation}
On the other hand, for $J_{\perp}>0$, 
$G={\rm colum}(-\frac{J_{\parallel}}{\gamma}2\pi a_n+n(J_{\perp}-h), nh,nh)$, which
leads to the form of the free energy (\ref{FE}) without the field term $h$. 
The TBA equations (\ref{TBA}) provide a clear  physical picture of the groundstate, 
the elementary excitations, as well as the thermodynamic quantities, such as the
free energy, magnetization, susceptibility, etc.
Our results extend the earlier calculations on this model \cite{ladd4, ladd2}.

\vskip 3mm
\noindent
{\bf II. Ferromagnetic rung coupling.} 
In the low-temperature regime $T\rightarrow 0$, only the negative part of the 
dressed energies $\epsilon^{(l)}$, denoted by $\epsilon^{(l)-}$, 
contribute to the ground-state energy. 
The TBA equations (\ref{TBA}) then become
\begin{eqnarray}
\epsilon^{(1)}&=&g_1-a_2*\epsilon^{(1)-}+a_1*\epsilon^{(2)-},\nonumber\\
\epsilon^{(2)}&=&g_2-a_2*\epsilon^{(2)-}+a_1*\left[\epsilon^{(1)-}+\epsilon^{(3)-}\right], \nonumber\\
\epsilon^{(3)}&=&g_3-a_2*\epsilon^{(3)-}+a_1*\epsilon^{(2)-},\label{TBA1}
\end{eqnarray}
where $g_a,\,a=1,2,3$, are the driving terms with respect to the basis order.
In the regime $J_{\perp}<0$, the component $|\!\! \uparrow \uparrow\rangle$ of the triplet state
is chosen as the reference state. The driving terms are given by 
$g_1=-\frac{J_{\parallel}}{\gamma} 2\pi a_1+h$, $g_2=h$ and $g_3=-h-J_{\perp}$, respectively.
Thus, in the absence of a magnetic field, the triplet is completely degenerate. 
The Fermi surface of the singlet is lifted as $J_{\perp}$ becomes more negative. 
If $J_{\perp}$ is negative enough, the singlet rung state is not 
involved in the groundstate, namely $\epsilon ^{(3)}(0)\geq 0$, 
whereas the two other triplet Fermi seas still have their Fermi boundaries at infinity. 
In such a configuration, we may determine the critical point defining the 
transition from the $su(4)$ phase into the $su(3)$ phase
by solving the TBA equations (\ref{TBA1}), with result
$J_c^{-}=-\frac{J_{\parallel}}{\gamma}(\frac{\pi}{\sqrt{3}}-\ln 3)$. 
At this critical point the free energy is given by 
$\frac{F(0,0)}{L}\approx -\frac{2J_{\parallel}}{3\gamma}(\psi(1)-\psi(\frac{1}{3}))$,
indicating a standard $su(3)$ phase. Here $\psi(n)$ is the digamma
function. It is worth noting that the critical point $J^{-}_c$ does
not stabilize if an external magnetic field is applied. If
the magnetic field is large enough, the ferromagnetic state
$|\!\!\uparrow\uparrow\rangle$ becomes the true physical groundstate,
i.e., there is a fully polarized gapped phase. It is found that for $h \geq
H_{c}^{{\rm F}}=\frac{4 J_{\parallel}}{\gamma}$, the state is
fully-polarized provided that $J_{\perp }\leq -\frac{4
J_{\parallel}}{\gamma}$. Therefore, in the ferromagnetic regime, 
the groundstate is in the critical $su(3)$ phase. If the magnetic field is 
greater than $H_c^{{\rm F}}$, the groundstate is ferromagnetic with a 
magnetization plateau $S^z=1$. 

\vskip 3mm
\noindent
{\bf III. Strong antiferromagnetic regime.}
In the antiferromagnetic regime, $J_{\perp}>0$, the rung singlet state is the reference
state. Thus the driving terms are given by 
 $g_1=-\frac{J_{\parallel}}{\gamma}2\pi a_1+J_{\perp}-h$ and $g_2=g_3=h$, respectively. 
From the TBA equations (\ref{TBA1}), if $h=0$ we immediately
conclude that the triplet excitation is massive with the gap
given by $\Delta =J_{\perp}-4J_{\parallel}/\gamma$
for the regime $J_{\perp}\geq
J^+_c=\frac{4J_{\parallel}}{\gamma}$. Here $J^+_c$ is the critical
point at which the quantum phase transition from the three branches of
the Luttinger liquid phase to the dimerized $u(1)$ phase occurs.  
To obtain good agreement with the experimental gap, we fix the rescaling constant
$\gamma$ with the coupling constants remaining arbitrary. For the strong coupling
compounds, e.g., ($5$IAP)$_2$CuBr$_4 \cdot 2$H$_2$O \cite{str3}, 
Cu$_2$(C$_5$H$_{12}$N$_2$)$_2$Cl$_4$ \cite{exp4},
(C$_5$H$_{12}$N)$_2$CuBr$_{4}$ \cite{exp5}, the experimental gap is
well established as $\Delta \approx J_{\perp}-J_{\parallel}$, and as a
consequence, we fix $\gamma \approx 4$.
On the other hand, for weak coupling compunds, e.g., (VO)$_2$P$_2$O$_7$ \cite{exp1,weak}, 
the choice $\gamma \approx 8$ determines a good fit with the gap
$\Delta \approx \frac{1}{2}J_{\perp}$. We stress that the 
purpose of introducing the rescaling
constant is to minimize the effects of the biquadratic term, so that the
model lies in the same Haldane phase as the pure Heisenberg ladder.


\vskip 3mm
\noindent
{\bf IV. Magnetization plateau.} The phase diagram of the
antiferromagnetic spin ladders in the presence of a magnetic field is
particularly interesting, because the critical points
can be measured through critical magnetic fields. The appearance of 
quantized magnetization plateaus in the presence of a strong magnetic field 
is expected on general grounds \cite{exp1}. From the TBA equations
(\ref{TBA1}) for antiferromagnetic rung coupling we observe that the
magnetic field lifts the Fermi seas of $\epsilon^{(2)}$ and
$\epsilon^{(3)}$. If $J_{\perp} >J_c^+$, we can show that the two components of 
the triplet states, $|3\rangle $ and $|4\rangle$, do not become involved in
the groundstate for a strong magnetic field. Basically, the magnetic field lifts
the component $|2\rangle$ of the triplet closer to the singlet groundstate 
such that they form a new effective spin-$\frac12$ state.
Therefore, in a strong magnetic field the groundstate may be considered
as a condensate of $su(2)$ hard-core bosons. The gap can be deduced via the
magnetic field $h$: the first critical field occurs at $H_{c1}$
where $g\mu_BH_{c1}=\Delta$, i.e. the magnetic field closes the
gap. The quantum phase transition from a gapped to a gappless
Luttinger phase occurs.  However, by continuing to increase the magnetic
field $h$ above the first critical field $H_{c1}$, the component $|2\rangle$ of the triplet 
becomes involved in the groundstate with a finite susceptibility. If the
magnetic field is greater than the rung coupling, i.e. $h>J_{\perp}$,
the state $|2\rangle$ becomes the lowest level. Therefore, it is
reasonable to choose the basis order as
$(|2\rangle,|1\rangle,|3\rangle,|4\rangle)^{{\rm T}}$. Subsequently
the driving terms are given by $g^{(1)}=-2\pi
J_{\parallel}a_1-J_{\perp}+h$, $g^{(2)}=J_{\perp}$ and
$g^{(3)}=h$. From the TBA, we see that the
groundstate is a fully-polarized ferromagnetic state
when the magnetic field is
greater than $H_{c2}=J_{\perp}+\frac{4J_{\parallel}}{\gamma}$.
Indeed,  the critical field $H_{c2}$ is in excellent agreement with the experimental data
for the very strong coupling compound ($5$IAP)$_2$CuBr$_4 \cdot 2$H$_2$O (abbreviated as B$5$i$2$aT),
\cite{str3}  and in a good agreement with the strong coupling compounds
 Cu$_2$(C$_5$H$_{12}$N$_2$)$_2$Cl$_4$  (abbreviated Cu(Hp)Cl) \cite{exp4} and
(C$_5$H$_{12}$N)$_2$CuBr$_{4}$ (abbreviated BPCN) \cite{exp5} (see Table 1). 
On the other hand, the precise structure of the compound
KCuCl$_3$ is not clear \cite{exp1}. It is
believed to exhibit a double-chain structure \cite{exp6} with a gap
$\Delta \approx 35K$ identified via the best fitting in the susceptibility
curve through the Troyer formula \cite{troyer}. The coupling constants are determined as
$J_{\perp}=4J_{\parallel},\,J_{\parallel}=12.3K,\, J_{{\rm diag}}=0$
\cite{exp6}. However, high-field measurements indicate the gap
$\Delta \approx 31.1K$ \cite{exp7}.  
Our TBA result gives poor agreement with the experimental result
for this type of ladder compound (see Table 1).
This suggests that the compound may exhibit a double-chain structure with additional
diagonal interaction. 
For these double-chain structure ladders, such as KCuCl$_3$, TlCuCl$_3$ etc., the leg
couplings appear to be very large, resulting in discrepancy with the
critical fields derived from the TBA method.

After a similar calculation, we obtain the magnetization 
$S^z\approx 4Q_1(1-2Q_1/\pi)/\pi$ in the vicinity  of the critical field $H_{c1}$, 
with the Fermi boundary $Q_1 \approx \sqrt{(h-H_{c1})/(H_{c1}-5h)}$. 
For a very strong magnetic field such that $H_{c2}-h<<1$ the free energy is
\begin{equation}
\frac{F(0,h)}{L}\approx -h -\frac{4}{\pi}\frac{(H_{c2}-h)^{\frac{3}{2}}}{\sqrt{5h-H_{c2}}},
\end{equation}
and the susceptibility $\kappa \approx \frac{3}{\pi \sqrt{4H_{c2}}}(H_{c2}-h)^{-\frac{1}{2}}$, 
which indicates the nature of the singular behavior in a phase
transition from a gapless to a ferromagnetic phase. The magnetization  is given by 
$S^z\approx 1-4Q_2(1-2Q_2/\pi)/\pi$, where $ Q_2  \approx \sqrt{(H_{c2}-h)/(5h-H_{c2})}$. 
The fact that the magnetization depends on the square root of the field in the vicinity of 
the critical fields is consistent with other theoretical \cite{FT1,FT2}
and numerical results \cite{numer1}. The magnetization increases almost
linearly between the critical field $H_{c1}$ and $H_{c2}$. The groundstate 
is ferromagnetic above $H_{c2}$ with the gap $\Delta =\mu g (H-H_{c2})$.

Numerical solution of the TBA equations gives 
a reasonable magnetization curve (see Fig. 1) which passes through an 
inflection point midway between $H_{c1}$ and  $H_{c2}$. 
This inflection point is clearly visible in experimental curves, e.g. for  
($5$IAP)$_2$CuBr$_4 \cdot 2$H$_2$O \cite{str3}, 
Cu$_2$(C$_5$H$_{12}$N$_2$)$_2$Cl$_4$ \cite{exp4} and 
(C$_5$H$_{12}$N)$_2$CuBr$_{4}$ \cite{exp5}.
The physical meaning of the inflection point is that the probabilities of the 
singlet and the  triplet states $|2 \rangle$ in the groundstate  are equal. 
It suggests an ordered dimer state close to half-filling \cite{IP}.
Therefore, in the strong-coupling regime, 
the one point correlation function $\langle S_j \cdot T_j\rangle =-\frac34$ 
lies in a gapped singlet groundstate, 
which indicates an ordered dimer phase, while 
$\langle S_j\cdot T_j\rangle =\frac14$ is in the fully-polarized ferromagnetic phase. 
However, in a Luttinger liquid phase, we find $\langle S_j\cdot T_j\rangle =-\frac34+S^z$. 
The magnetic field increases the one point correlation function. 

We also notice that our results for the gap, $\Delta =J_{\perp}-4J_{\parallel}/\gamma$,  
and the critical field, $H_{c2}=J_{\perp}+4J_{\parallel}/\gamma$,  
coincide for $\gamma =4$ with the first-order perturbation theory
results obtained for strong coupling \cite{FT1}. 
However, their higher-order terms lead to poor agreement with the experimental
results. It is apparent that the rescaling constant $\gamma$ causes a shift in the critical point. 
This can be seen from the values of
\begin{equation} 
H_{c2}/\Delta = 1+2/(\frac{\gamma J_{\perp}}{4J_{\parallel}}-1),
\label{crit}
\end{equation}
which are plotted in Fig. 2.
The larger the ratio of $\frac{J_{\perp}}{J_{\parallel}}$, 
the closer the two critical points are.
This means that the critical points $H_{c1}$ and  $H_{c2}$ cannot be distinguished
for a very large energy gap. Once the gap is closed by an external field, 
the groundstate immediately becomes fully-polarized. This is evident in 
the strong coupling compound ($5$IAP)$_2$CuBr$_4 \cdot 2$H$_2$O \cite{str3}. 
Here the gap opens only if $J_{\perp}/J_{\parallel}\geq 4/\gamma$, with $\gamma$ arbitrary.
And therefore the gap opens for arbitrary $ J_{\perp}/J_{\parallel}$.

\begin{center}
\begin{table}
\begin{tabular}{|c|c|c|c|c|c|c|c|c|}
\hline 
Compounds&$g$ & $J_{\perp }$ & $J_{\parallel}$ & $\gamma$ &$H_{c1}$(exp)& 
$H_{c2}$(exp) &$H_{c1}$(TBA) &$H_{c2}$(TBA)\\
\hline 
  B$5$i$2$aT
&2.1 &
$13K$&
$1.15K$& 
$4$&
$8.4T$&
$10.4T$ & 
$8.3T$ &
$ 10.03T$
\\
\hline
 Cu(Hp)Cl  &
2.03&
$13.2K$&
$2.5K$& 
$4$&
$7.5T$&
$13.2T$ & 
$7.84T$ &
$11.51T$
\\
\hline 
BPCB 
&2.13&
$13.3K$&
$3.8K$& 
$4$&
$6.6T$&
$14.6T$ & 
$6.6T$ &
$11.95T$
\\
\hline 
KCuCl$_3$
&2.05&
$49.2K$&
$12.3K$& 
$2.68$&
$22.4T$&
$\approx 60T$ & 
$22.4T$ &
$ 49T$
\\
\hline 
\end{tabular}
\caption{Comparison between the experimental values for the critical points 
$H_{c1}$ and $H_{c2}$ for strong coupling ladder compounds and the TBA 
results obtained from the $su(4)$ integrable model.} 
\end{table}
\end{center}

\begin{figure}
\centerline{\epsfig{file=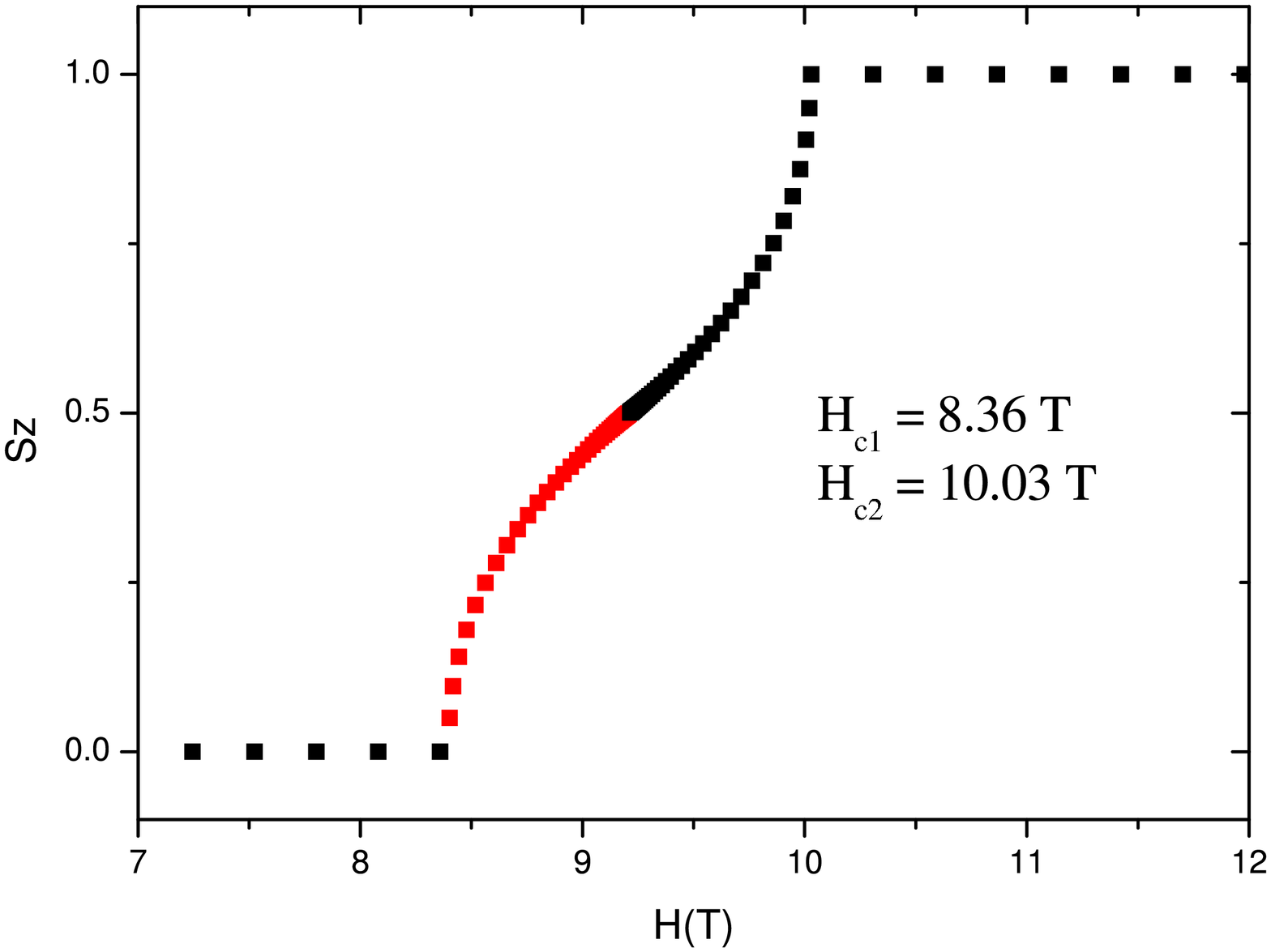,width=8cm,height=11cm,angle=-90}}
\caption{Magnetization $S^z$ versus magnetic field $H=\mu_Bgh$ obtained
from the TBA equations for the values $J_{\perp}=13K$, $J_{\parallel}=1.15K$ and $\gamma=4$
for the strong coupling compound ($5$IAP)$_2$CuBr$_4 \cdot 2$H$_2$O \cite{str3}.
At the inflection point $h=J_{\perp}$ the magnetization is $0.5$. 
The curve is in excellent agreement with the experimental result \cite{str3}.}
\end{figure}

\begin{figure}
\centerline{\epsfig{file=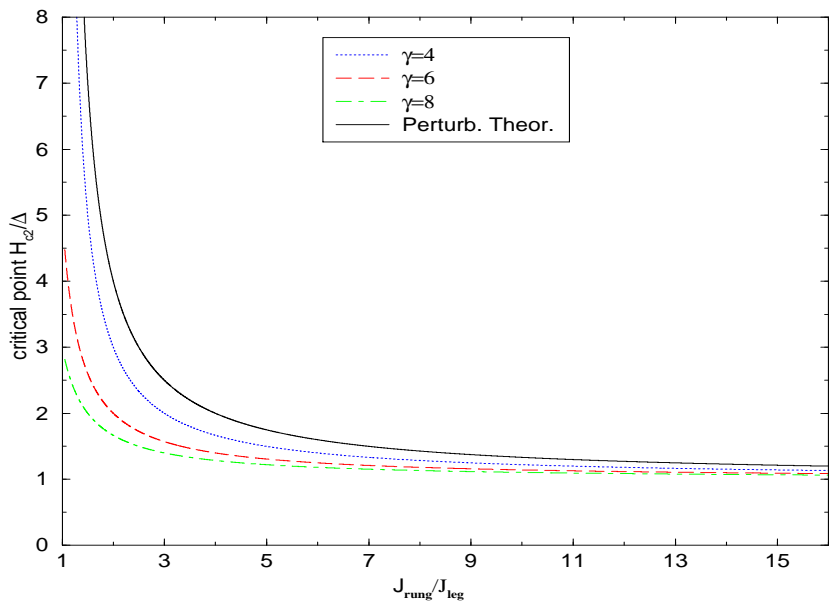,width=8cm,height=11cm,angle=-90}}
\caption{The critical point ${H_{c2}}/{\Delta}$ as a function 
of the ratio $J_{\perp}/J_{\parallel}$ for different values of the rescaling 
parameter $\gamma$. Also shown is the perturbation theory result.} 
\end{figure}

Finally, we show the phase diagram in the presence of a magnetic field in Fig. 3. 
In the ferromagnetic rung coupling regime, the fully-polarized ferromagnetic state 
lies in the region $h\geq |J_{\perp}|$ and $h\geq 4J_{\parallel}/\gamma$, 
whereas the $su(3)$ Luttinger magnetic phase is in the region $h<|J_{\perp}|$ 
and left of the boundary between the $su(3)$ and $su(4)$ phases. 
The $su(4)$ phase is in the region $h< 4J_{\parallel}/\gamma$ and right of this boundary. 
In the antiferromagnetic rung coupling regime, the singlet rung state lies in  
$h<J_{\perp}-4J_{\parallel}/\gamma$ whereas the ferromagnetic fully-polarized state 
is in the region $h\geq J_{\perp}+4J_{\parallel}/\gamma$. The $su(2)$ magnetic phase 
remains in the region $h> J_{\perp}-4J_{\parallel}/\gamma$, 
$h <J_{\perp}+4J_{\parallel}/\gamma$ and $J_{\perp}\geq 4J_{\parallel}/\gamma$. 
The $su(4)$ magnetic  phase  lies in the region $h<J_{\perp}+4J_{\parallel}/\gamma$ 
and $0<J_{\perp}<4J_{\parallel}/\gamma$.

\begin{figure}
\centerline{\epsfig{file=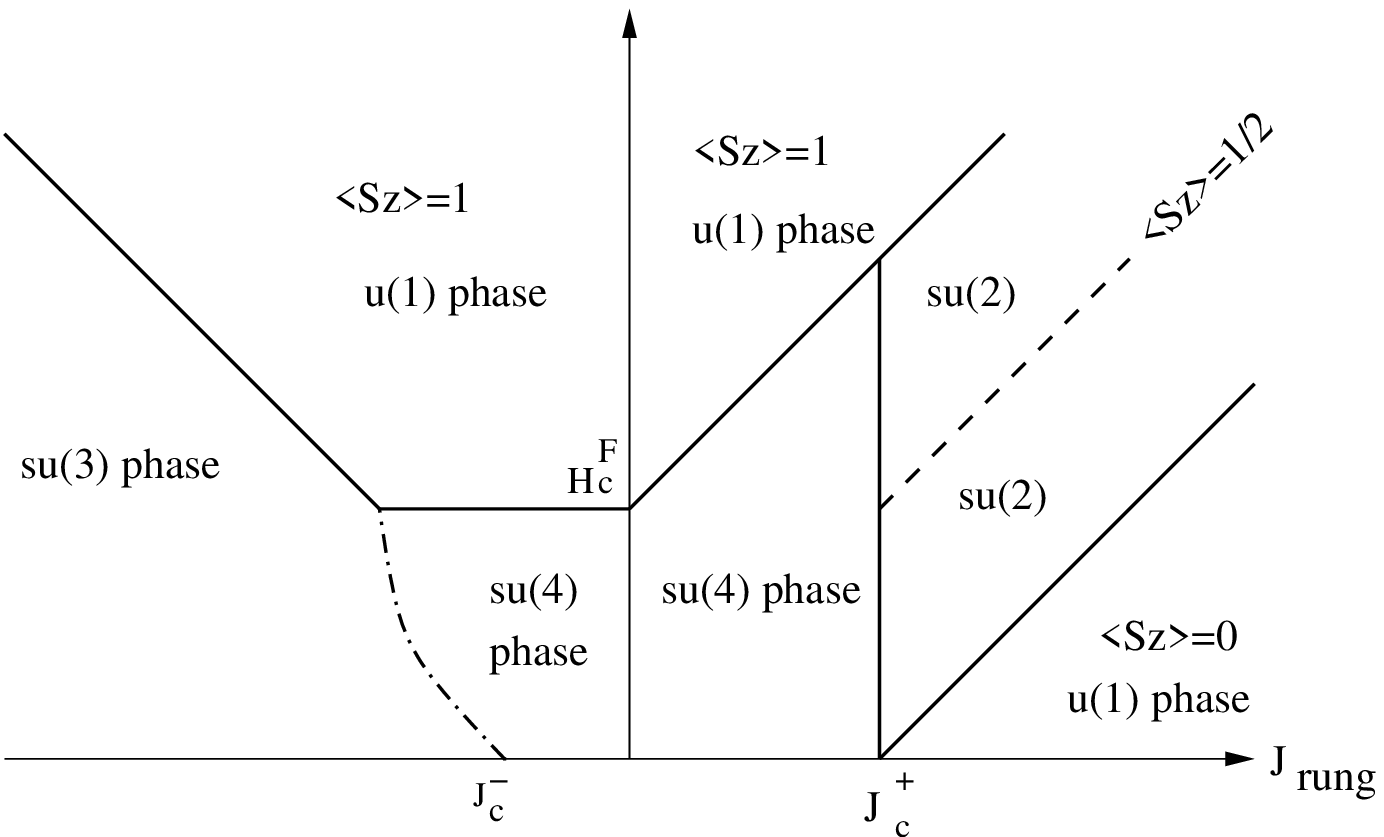,width=11cm,height=8cm,angle=0}}
\caption{The magnetic phase diagram of the two-leg $su(4)$ ladder.
In the antiferromagnetic regime the thick lines are $h= J_{\perp}-4J_{\parallel}/\gamma$, 
$h= J_{\perp}+4J_{\parallel}/\gamma$ and the dash line is  $h=J_{\perp}$. 
In the ferromagnetic regime the thick lines are $h=-J_{\perp}$ and $h=4J_{\parallel}/\gamma$. 
The dash-dot line is an approximate boundary between the $su(4)$ and $su(3)$ phases. }
\end{figure}

To conclude, we have studied the phase diagram of the integrable $su(4)$ spin
ladder model (1) by means of the TBA. 
In particular, the critical behavior at the critical points  
$H_{c1}$ and $H_{c2}$ was derived. In the presence of a strong magnetic field, the phase 
diagram is in a good agreement with the experimental observations for the strong 
coupling compounds 
($5$IAP)$_2$CuBr$_4 \cdot 2$H$_2$O \cite{str3}, Cu$_2$(C$_5$H$_{12}$N$_2$)$_2$Cl$_4$ \cite{exp4}
and (C$_5$H$_{12}$N)$_2$CuBr$_{4}$ \cite{exp5}.
We have also predicted the spin gap $\Delta \approx J_{\perp}-\frac{1}{2}J_{\parallel}$ 
for the weak coupling compounds with $ J_{\perp} \sim J_{\parallel}$, such as 
$(VO)_2P_2O_7$ and also shown that the gap opens for arbitrary $J_{\perp}/ J_{\parallel}$.

\ack
The authors thank the Australian Research Council for support. 
A.F. and X.W.G. thank I. Roditi and Z.-J. Ying for helpful discussions as well as FAPERGS  
for financial support.


\vspace{1cm}
 
\begin{thebibliography}{99}
%
\bibitem{exp1} Dagotto E and Rice T M 1996 Science {\bf 271} 618

Dagotto E 1999 Rep. Prog. Phys. {\bf 62}  1525
%
\bibitem{exp2} Azuma M, Hiroi Z, Takano M, Ishida K and Kitaoka Y 1994 
Phys. Rev. Lett. {\bf 73}  3463
%
\bibitem{exp4} Chaboussant G \etal 1997 Phys. Rev. Lett. {\bf 79}  925

Chaboussant G \etal 1998 Phys. Rev. Lett. {\bf 80}  2713

Chaboussant G \etal 1997 Phys. Rev. {\bf B55}  3046
%
\bibitem{exp5} Watson B C, Kotov V N and Meisel M W 2001 
Phys. Rev. Lett. {\bf 86}  5168
%
\bibitem{str3} Landee C P, Turnbull M M, Galeriu C, Giantsidis J and  Woodard F M 
2001 Phys. Rev. {\bf B63} 100402
%
\bibitem{exp6} Tanaka H, Takatsu K, Shiramura W and Ono T 1996 
J. Phys. Soc. Japan {\bf 65} 1945

Nakamura T and Okamoto K 1998  Phys. Rev. {\bf B58}  2411
%
\bibitem{exp7} Shiramura W \etal 1997 J. Phys. Soc. Japan  {\bf 66}  1900
%
\bibitem{ladd1} Nersesyan A A and Tsvelik A M 1997 Phys. Rev. Lett. {\bf 78} 3939

Kolezhuk A K and Mikeska H-J  1998 Phys. Rev. Lett. {\bf 80} 2709
%
\bibitem{numer1} Zheng W, Singh R R P and Oitmaa J 1997 Phys. Rev. {\bf B55}  8052
%
\bibitem{numer2} Dagotto E, Riera J and Scalapino D 1992
Phys. Rev. {\bf B45} 5744
%
\bibitem{FT1} Reigrotzki M, Tsunetsugu H and Rice T M 1994 
J. Phys. Condens. Matter {\bf 6} 9235

Giamarchi T and Tsvelik A M 1999 Phys. Rev. {\bf B59} 11398
%
\bibitem{FT2}
Totsuka K 1998  Phys. Rev.{\bf  B57} 3454
%
Barnes T and Riera J 1994 Phys. Rev. {\bf B50} 6817.
%
\bibitem{ladd4} Wang Y 1999 Phys. Rev. {\bf B60} 9236

\bibitem{ladd3} Batchelor M T and Maslen M 1999 J. Phys. A  {\bf 32} L377 

Frahm H and Kundu A 1999 J. Phys. C: Cond. Mat. {\bf 11}  L557

de Gier J, Batchelor M T and Maslen M 2000  Phys. Rev. {\bf B61} 15196

Batchelor M T, de Gier J and Maslen M 2001 J Stat. Phys. {\bf 102}  559

Maslen M, Batchelor M T and de Gier J, cond-mat/0302135.
%
\bibitem{ladd5} Park S and Lee K 1998 J. Phys. A: Math. Gen. {\bf 31} 6569

\bibitem{weak} Johnston D C, Johnson J W, Goshorn D P and Jacobson A J 1987
Phys. Rev.{\bf  B35} 219

\bibitem{Hayw} Hayward C A and Poilblanc D 1996  Phys. Rev. {\bf B54} R12649

\bibitem{BA} Sutherland B 1975 Phys. Rev. {\bf B12} 3795

Schlottmann P 1992 Phys. Rev. {\bf B45}  5293
%
\bibitem{TBA} M. Takahashi 1971  Prog. Theor. Phys. {\bf 46}  401

P. Schlottmann 1986  Phys. Rev. {\bf B33}  4880.
%
\bibitem{Lee}K. Lee 1994 J. of the  Korean Phys. Soc. {\bf  27}  205.

\bibitem{ladd2} de Gier J and Batchelor M T 2000 Phys. Rev. {\bf B62 } R3584

Cai S, Dai J and Wang Y 2002 Phys. Rev. {\bf B66} 134403

\bibitem{troyer} Troyer M, Tsunetsugu H and W\"{u}rtz D 1994 Phys. Rev. {\bf B50} 13515

\bibitem{IP} Chaboussant G \etal 1998 Eur. Phys. J. B {\bf 6} 167 


\end{thebibliography}
\end{document}